\documentclass[preprint]{aastex701}

\begin{document}

\title{Near-Discovery Observations of Interstellar Comet 3I/ATLAS\\with the NASA Infrared Telescope Facility}

\author[orcid=0000-0003-1008-7499]{Theodore Kareta}
\affiliation{Department of Astrophysics and Planetary Science, Villanova University, Villanova, PA, USA}
\affiliation{Lowell Observatory, Flagstaff, AZ, USA}
\email[show]{theodore.kareta@villanova.edu}

\author[0009-0001-1093-3715]{Chansey Champagne}
\affiliation{Department of Astronomy and Planetary Science, Northern Arizona University, Flagstaff, AZ, USA}
\email{cc3776@nau.edu}

\author[0000-0002-8041-0573]{Lucas McClure}
\affiliation{Department of Astronomy and Planetary Science, Northern Arizona University, Flagstaff, AZ, USA}
\email{tm87@nau.edu}

\author[0000-0001-9265-9475]{Joshua Emery}
\affiliation{Department of Astronomy and Planetary Science, Northern Arizona University, Flagstaff, AZ, USA}
\email{joshua.emery@nau.edu}

\author[0000-0003-1383-1578]{Benjamin N.L. Sharkey}
\affiliation{Department of Astronomy, University of Maryland, College Park, MD, USA}

\email{sharkey@umd.edu}
\author[0000-0001-9542-0953]{James Bauer}
\affiliation{Department of Astronomy, University of Maryland, College Park, MD, USA}
\email{gerbsb@umd.edu}

\author{Michael S. Connelley}
\affiliation{Institute for Astronomy, University of Hawaii at Manoa, Manoa, HI, USA}
\email{msconnelley@gmail.com}

\author{John Rayner}
\affiliation{Institute for Astronomy, University of Hawaii at Manoa, Manoa, HI, USA}
\email{john.thornton.rayner@gmail.com}

\author[0000-0003-3091-5757]{Cristina A. Thomas}
\affiliation{Department of Astronomy and Planetary Science, Northern Arizona University, Flagstaff, AZ, USA}
\email{cristina.thomas@nau.edu}

\author{Vishnu Reddy}
\affiliation{Lunar and Planetary Laboratory, University of Arizona, Tucson, AZ, USA}
\email{vishnureddy@arizona.edu}

\author[0009-0007-6497-5156]{Megan Firgard}
\affiliation{Department of Physics, University of Central Florida, Orlando, FL, USA}
\affiliation{Florida Space Institute, University of Central Florida, Orlando, FL, USA}
\email{megan.firgard@gmail.com}

\begin{abstract}
Interstellar Objects are comets and asteroids that formed around other stars but were ejected before they could accrete into exoplanets. They therefore represent a rare opportunity to compare the the building blocks of planets in the Solar System to those in other stellar systems. The third Interstellar Object, 3I/ATLAS, is the newest, brightest, potentially largest, and fastest member of this population. We report observations of 3I/ATLAS taken on 2025 July 3 and 4 with the NASA Infrared Telescope Facility just days after its discovery. In $r'$-band imaging with 'Opihi, we see no obvious lightcurve variability and derive a $g'-i'$ color of $0.98\pm0.03$ which is consistent in spectral slope to other near-discovery observations. We obtained the first near-infrared (NIR) reflectance spectrum of 3I/ATLAS with SpeX. The visible color and NIR spectrum show a linear, red visible slope, a somewhat less red slope between 0.7 and 1.1 $\mu$m, and a neutral or slightly blue slope at longer wavelengths. Challenges in modeling the reflectivity of 3I may indicate that this comet has a complex grain size distribution, grain compositions unlike Solar System comets, or both. Like 2I/Borisov, there are no obvious signatures of water ice in the coma of 3I/ATLAS. Observations closer to perihelion will help elucidate whether 3I has less water than anticipated or whether the Interstellar Objects might retain and release their ices somewhat differently than Solar System comets.
\end{abstract}

\section{Introduction} \label{sec:intro}
The Interstellar Objects (ISOs) are small bodies analogous to our own comets and asteroids which formed around other stars, were subsequently ejected, and then were discovered passing through the Solar System on hyperbolic trajectories. The small bodies of the Solar System are its remnant planetsimals -- the small portion of the planetesimals that did not get accreted to form the larger planets, get ejected, or fall into the Sun -- and we expect that the same is true for the ISOs. Studying these objects individually and as a population therefore provides a means to probe the specifics of planet formation on a galactic scale. Two primary challenges encountered thus far have been the small number of objects discovered (before 2025, only two) and interpreting the properties of both of those objects due to their difference from Solar System objects (see \citealt{2023ARA&A..61..197J} for a recent review).

The third ISO discovered, C/2025 N1 (ATLAS), now designated 3I/ATLAS, was found on 2025 July 1 by the ATLAS project's \citep{2018PASP..130f4505T} telescope in Chile (see a review of its discovery circumstances in \citealt{2025arXiv250702757S}). At discovery, the object was approximately $m_V\sim18$ in brightness and slightly diffuse, indicating that it was likely undergoing some kind of cometary mass loss. Given that 3I was discovered at opposition (phase angle $\alpha\approx2^{\circ}$), any potential tail would have been pointed away from the Earth and partially obscured by the foreground coma. 3I's eccentricity ($e=6.13\pm0.02$) is significantly higher than the previous two ISOs and may be a good indication that it samples a different part of the galaxy \citep{2025arXiv250705318H} when taken with other aspects of its orbit. It is also intrinsically brighter than the previous two ISOs. Its bright absolute magnitude of $H_V\approx12$ cannot be directly converted into a diameter given the likely substantial coma contribution, but does constrain it to be at most $20$ km across \citep{2025arXiv250702757S}.

Spectral and photometric observations reported in \citet{2025arXiv250702757S} show that 3I's reflectance spectrum at visible wavelengths is approximately linear (without spectral curvature) and relatively red ($S'=17.1\pm0.2\%$), comparable to the surfaces of comets and D-type asteroids (see, e.g., \citealt{2009Icar..202..160D}). Precovery observations suggest that the object may have been more neutral in color prior to its July 1 discovery \citep{2025arXiv250702757S}. Observations from the Very Large Telescope (VLT) reported in \citet{2025arXiv250705226O} also show a red and linear visible slope ($S'=18\pm4\%$), the Gran Telescopio Canarias (GTC) was used to measure $S'=18.3\%$ \citep{2025arXiv250712922D}, and photometric observations from the Palomar 200" show similar slopes from multi-filter photometry \citep{2025arXiv250705252B} and spectroscopy \citep{2025RNAAS...9..194B}.

Notably, despite higher spectral resolution and a larger aperture telescope, no gases were spotted in the vicinity of the comet in the VLT dataset (\citealt{2025arXiv250705226O}, see also \citealt{2025arXiv250707312A}) or GTC dataset \citep{2025arXiv250712922D}. This contrasts with the earliest observations of 2I/Borisov (see, e.g, \citealt{2019ApJ...885L...9F,2019A&A...631L...8O,2020ApJ...889L..38K, 2020MNRAS.495.2053D}) in which CN gas was detected rapidly with ground-based optical telescopes and then followed by other gases in quick succession (see, e.g., \citealt{2020ApJ...889L..30L,2020ApJ...889L..10M}). The absence of gas around the time of discovery might be due to intrinsic differences between 3I and 2I, such as the predicted water-dominated composition of 3I suggested by the modeling of \citet{2025arXiv250705318H}. 3I/ATLAS was, however, discovered nearly $4.5$ au from the Sun compared to approximately $3.0$ au for Borisov, and thus would be expected to be colder and less active early on. Although the larger heliocentric distance at which 3I was discovered might be an impediment to early detection of gas species, it might aid in the detection of water ice in its coma due to the longer time it takes ice to sublimate at those distances. Despite Borisov showing clear signs of water vapor production (see, e.g., \citealt{2020ApJ...893L..48X}), no clear signs of water ice were spotted in its coma (see, e.g., \citealt{2020A&A...634L...6Y,2020AJ....160..132L}).

In this paper, we present and analyze observations from SpeX and `Opihi on the NASA Infrared Telescope Facility (IRTF) taken within three days of 3I/ATLAS's discovery to measure its reflectivity and search for signs of water ice. The early nature of our observations should allow more direct comparison with other observations (e.g., those in the same activity state) in the discovery epoch than those taken later.

\section{Observations} \label{sec:obs}
We observed 3I/ATLAS on UTC 2025 July 3 and 4 with SpeX (\citealt{2003PASP..115..362R}, later upgraded in 2014) on the IRTF utilizing previously-scheduled (e.g., non-discretionary) time. On July 3 (July 4), 3I was $R_H=4.43$ ($4.40$) au away from the Sun, $\Delta=3.43$ ($3.40$) au away from the Earth, and at a phase angle of $\alpha=2.6^{\circ}$ ($2.9^{\circ}$). SpeX was configured in the low-resolution `prism' mode with a slit-width of $0.8$", resulting in an effective resolution of approximately $R\sim70$ increasing to $R\sim150$ over the $0.7-2.5\mu{m}$ wavelength range. Given that the object was only expected to be $1.5-2.0$" in diameter, the 15"-long slit was used in a standard ABBA nodding pattern with the slit aligned along the parallactic angle ($\pm 10^{\circ}$). Acquisition and guiding was accomplished with the visible-wavelength guider camera MORIS \citep{2010DPS....42.4914G}. We also utilized the wide-field acquisition camera `Opihi \citep{2022SPIE12184E..8DL} with the Sloan $r'$ filter on July 3 and alternating between the $g'$ and $i'$ filters on July 4 to obtain lightcurves and constrain 3I's visible reflectivity. Given that time-sequence color photometry was being performed with `Opihi, filters were chosen for MORIS to try to maximize reliability of guiding as opposed to obtain visible colors or a lightcurve as is sometimes done in small body observations. Due to the object's proximity to the galactic plane, acquisition of the object into the SpeX slit was challenging, and only frames in which we could be sure there was minimal contamination were reduced and analyzed further, corresponding to twenty frames on July 3 and fifty frames on July 4.  The on-chip integration time was 120 s per frame on both nights. Atmospheric conditions (i.e., seeing) were significantly better on July 4, which, combined with the larger number of frames, led to better S/N for the spectrum that night. A local G3V standard star, HD164554, was observed at multiple airmasses both nights. The well-characterized solar analog Landoldt SA 107-684 was observed on July 3 and BS4486 on July 4.
 
Simultaneous $r'$ (July 3) and $g', i'$ (July 4) imaging data were obtained using the IRTF's 'Opihi facility camera \citep{2022SPIE12184E..8DL}. Attached to the IRTF's 17-inch finder scope, the 2048$\times$2048 CCD camera with 0.94 arcsecond pixels yields images with a $0.5\times0.5$ square degrees field of view. (These large pixels made the extension of the target less obvious, but efforts to stack the 'Opihi and MORIS data will be pursued in more depth in a later paper.) Sixty-second exposures were taken between dithers for the SpeX observations. Frames that were concurrent with telescope dithers were rejected, as were frames where the ISO was co-located on the sky with background objects of similar or greater brightness. On the night of July 3, the 'Opihi imaging yielded 129 $r'$ band exposures spanning an interval of approximately 3 hours and 40 minutes. The following night of July 4 72 $g'$ band and 43 $i'$ band images of sufficient quality were obtained over a timespan of 3 hours and 55 minutes.

Reduction of the SpeX data followed the procedure described in \citet{Emery2011}.  Briefly, all frames were divided by a flat field (measured using an integrating sphere attached to Spex) and background frames were subtracted.  Normally, one subtracts A-B pairs as a first-order sky emission removal, but the field was so crowded that many pairs contained stars in the background nod position.  We therefore created median sky frames for each nod position each night using non-contaminated frames. The object position was located and summed within a 6-pixel-wide aperture.  Standard stars were extracted similarly (though using normal A-B subtraction), and the object frames divided by star frames measured close in airmass. The resulting reflectance frames were normalized and averaged to compute the final spectrum each night.  Wavelength calibration was applied using spectra of an Argon lamp and extracted using \textit{Spextool} \citep{2004PASP..116..362C}. A portion of the July 4 dataset was reduced by a second co-author fully within the \textit{Spextool} environment as a consistency check.

Reduction of the 'Opihi data was done using a combination of IRAF routines \citep{1986SPIE..627..733T}, and astropy python libraries \citep{2022ApJ...935..167A}. Per-frame astrometric solutions were computed using astrometry.net \citep{2010AJ....139.1782L}. Zero points for each frame were computed by matching tens of field stars to the PANSTARRS1 catalog \citep{panstarrs2016} using \textit{calviacat} \citep{calviacat2021}, and thus reported magnitudes are within the PANSTARRS system. Aperture photometry was extracted using 4-pixel radius circular apertures on the object for all filters on both nights.

\section{Results} \label{sec:results}
\subsection{Visible Photometry}
\begin{figure*}[ht!]
\plotone{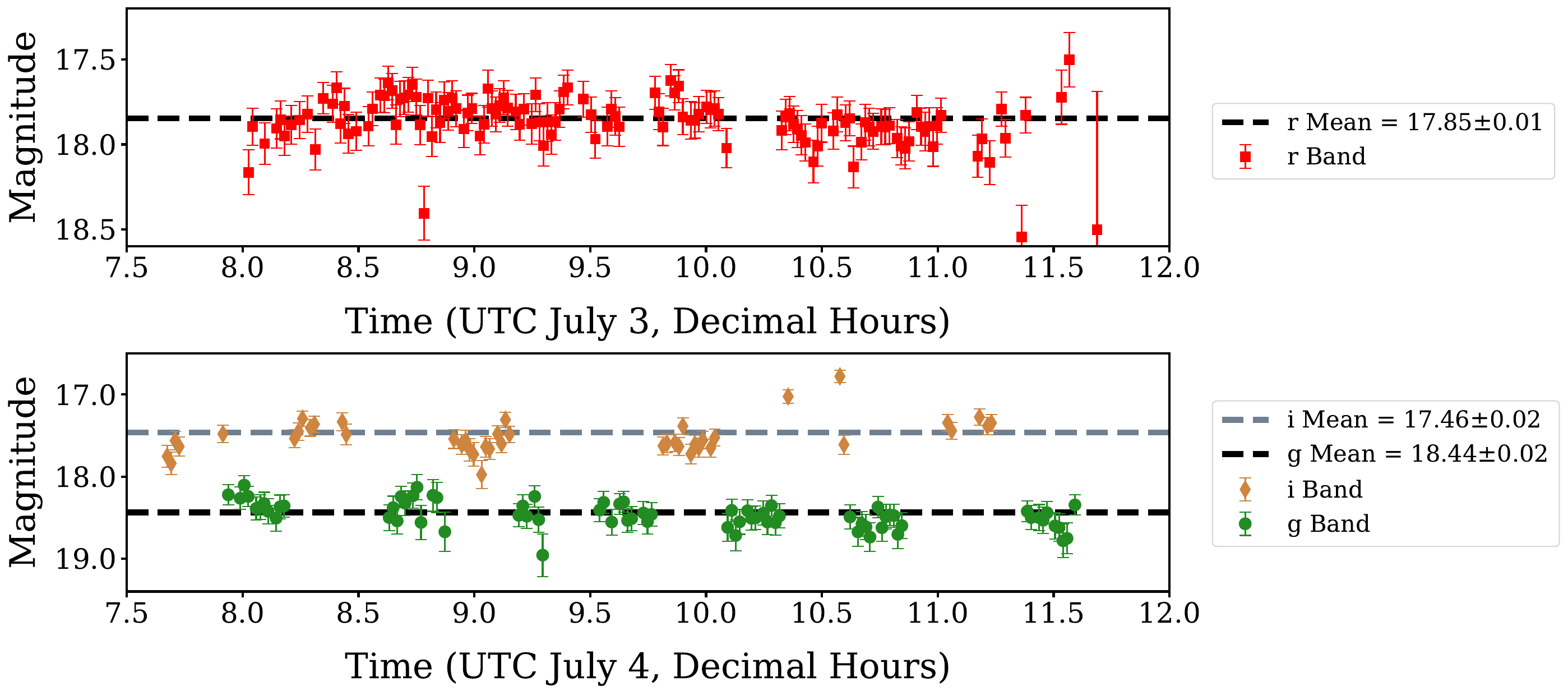}
\caption{Top: Slightly more than three hours of Sloan $r'$ filter observations of 3I/ATLAS from 'Opihi \citep{2022SPIE12184E..8DL} on 2025 July 3. Bottom: Interwoven $g'$ and $'i$ lightcurves from July 4. The lightcurve of 3I is primarily flat or stable, with variation roughly on the scale of the uncertainty in each datapoint, and at least some of the scatter may be from faint background stars in the crowded field. The visible colors derived from the best frames on July 4 support a red and linear visible reflectivity similar to what has been measured by other observers, see text for more details.
\label{fig:lc}}
\end{figure*}

We present our 2025 July 3 Sloan $r'$ lightcurve from 'Opihi in Figure \ref{fig:lc}. The lightcurve shows little variation over the three consecutive hours of observations. The scatter in the individual datapoints is barely larger than the error bars on individual datapoints themselves. 3I was discovered within the galactic plane, and performing uncontaminated photometry of it has proved challenging for us and other observers. We discarded any frame where 3I was obviously contaminated by a nearby star, but faint background sources still must contribute to some of the scatter seen in Figure \ref{fig:lc}. Therefore, we do not interpret our lightcurve as showing any obvious rotation-based signal. Although the absence of rotational variation is consistent with the object being spherical or at least circular along the line of sight, it is more likely that any rotational signal is masked or muted by the dust in the object's coma. If the object has a large nucleus and relatively weak activity, its nuclear lightcurve might be more apparent in the coming months at higher phase angles. If 3I is closer in size to a typical Solar System comet, we may not be able to measure its lightcurve until after its activity ceases on the way outbound -- if it's still bright enough to do so.

If we take the average apparent magnitude measured in Sloan $r'$ over the three hours on July 3 and correct for geometric and heliocentric distances, we retrieve a phase-dependent absolute magnitude of $H_V(\alpha)=12.10\pm0.01$. Although it is not clear what 3I's actual phase curve looks like yet (and if its activity state is changing, it may be hard to determine for some time), typical cometary phase curves have linear coefficients of $\beta=0.04$ magnitudes per degree \citep{2017MNRAS.471.2974K}. Using this average phase coefficient for 3I, its true $\alpha=0^{\circ}$ absolute magnitude would be $11.98\pm0.01$. These results agree with measurements by other observers (see, e.g., a review of $H_V$ values in \citealt{2025arXiv250721967F}).

The following night $g'$ and $i'$ lightcurves were extracted from the 115  'Opihi select frames taken on July 4 in which the target was very well separated from nearby stars. We derive an average $g'-i'$ color of $0.98\pm0.03$ from the full observational set, including only images where 3I/ATLAS was not overlapping any identifiable background field star. As can be seen in Figure \ref{fig:spectra}, this results in a visible slope quite similar to the \citet{2025arXiv250702757S} spectrum, and thus the \citet{2025arXiv250705226O, 2025arXiv250705252B, 2025arXiv250712922D} spectra as well which report very similar slopes. \citet{2025arXiv250705252B} reports a $g'-i'=1.00\pm0.05$ color, which agrees with our measured color closer to less than one standard deviation. 
We do note that lightcurve effects and contamination from faint background sources may still be present, which could systematically alter the nightly-averaged color measurements we present here. Such effects are difficult to quantify given our data set, but we note that both the $r'$ and $g'$ data sets vary by $\sim 0.1$ magnitudes over the course of July 3 and July 4, respectively. We do not detect any clear trends in $i'$ values, though this data was the most severely affected by background sources. Notably, \citet{2025arXiv250712922D} report a rotation period of $P=16.79$ hours for 3I/ATLAS based on near-discovery photometry (e.g., from near in time to ours). Our lightcurves strongly demonstrate limited ($<0.1$ magnitude) variability on short timescales, but this does not preclude somewhat larger variation like what they report ($\approx0.3$ magnitudes, \citet{2025arXiv250712922D}) so long as it is over this significantly longer timescale. While 3I/ATLAS has brightened somewhat since discovery making further attempts to measure its rotational period more challenging, it could be of interest to attempt to remeasure its nuclear lightcurve as it recedes from the Sun in 2026.

\subsection{Near-Infrared Spectroscopy}
\begin{figure*}[ht!]
\plotone{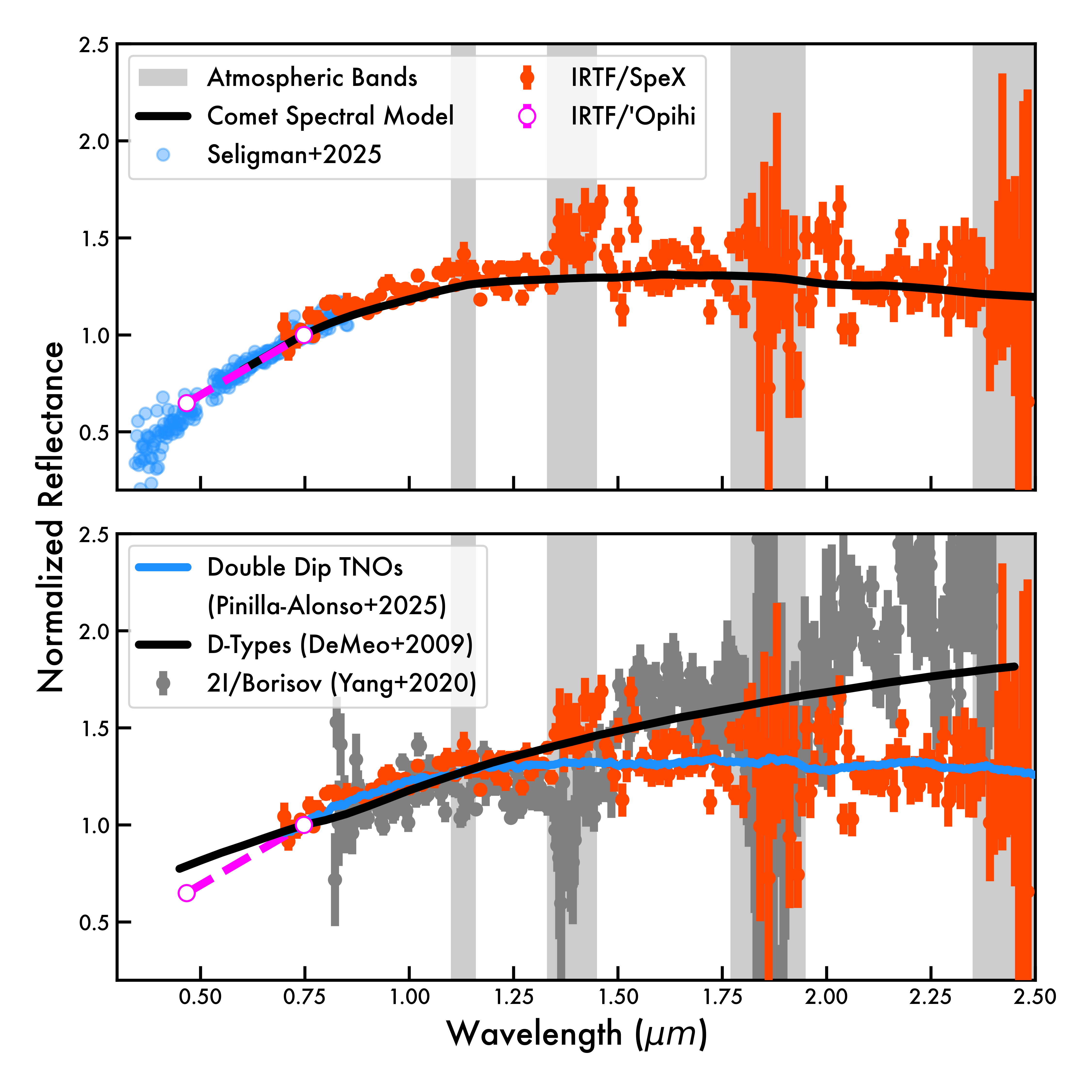}
\caption{Our SpeX spectrum from 2025 July 4 is shown in red and our 'Opihi $g'-i'$ colors converted to reflectance with the Solar colors of \citet{2018ApJS..236...47W} as magenta unfilled circles. All datasets and models are normalized at the wavelength of Sloan $i'$ ($0.748\mu{m}$) for convenience. Top: the visible reflectance spectrum of \citet{2025arXiv250702757S} is shown in blue for context and comparison to validate our spectral reductions. In the region where the two spectra overlap ($0.7\mu{m}-0.85\mu{m}$), the slopes agree well. Our maximum likelihood spectral model is shown in black. Bottom: Several spectral comparison to our 3I spectrum are shown, including the average Double Dip TNO from \citet{2025NatAs...9..230P}, and the average D-type asteroid from \citet{2009Icar..202..160D}, and a near-infrared spectrum of 2I/Borisov from \citet{2020A&A...634L...6Y}. Areas of significant atmospheric absorption (and thus significantly lowered throughput) are shaded in gray in both frames.
No obvious absorption signatures are seen in the reflectance spectrum of 3I/ATLAS.
\label{fig:spectra}}
\end{figure*}

We present the NIR reflectance spectrum of the 3I taken on July 4 in Figure \ref{fig:spectra}. The $g'-i'$ color derived from `Opihi are also included, as well as the visible spectrum of 3I from \citet{2025arXiv250702757S} for context. These and other datasets (described below) were all normalized at the wavelength of the Sloan $i'$ ($0.748\mu{m}$). The NIR spectrum from July 3rd has significantly lower S/N, but agrees in spectral slope and shape, indicating no substantial change in the spectral properties of the coma over the two nights. As can be seen, the agreement between SpeX, `Opihi, and the visible spectrum from \citet{2025arXiv250702757S} at the overlap wavelengths is quite good. The color derived from our best `Opihi frames, $g'-i'=0.98\pm0.03$, is consistent with the visible slopes of \citet{2025arXiv250702757S} and \citet{2025arXiv250705226O} as well as the colors of \citet{2025arXiv250705252B} ($g'-'i=1.00\pm0.05$). Although the short end of the NIR spectrum below $1.1\mu{m}$ is red and linear like most of the slopes reported for 3I at short wavelengths, the reflectance spectrum begins to decrease in slope beyond this point and approaches neutral reflectivity by $\sim1.25\mu{m}$. Beyond approximately $1.5\mu{m}$, the spectrum is consistent with flat within errors. The amount of spectral curvature, especially considering the long-wavelength negative slopes, clearly differentiates the object from the D-type asteroids it appeared similar to with visible wavelength data only. While no asteroid type matches 3I's spectrum well, the Double-Dip TNOs \citep{2025NatAs...9..230P} are quite similar in reflectivity. The D-type asteroids, the Double-Dip TNOs, and our comet model from Section \ref{sec:models} are also shown in Figure \ref{fig:spectra}. However, at least some of the light we see is reflected from the object's coma as opposed to from a solid surface, so the comparison between the asteroids, TNOs, or inactive other Solar System bodies may be of limited reliability.

\section{Ice Abundance Modeling} \label{sec:models}
In order to estimate the maximum allowable water ice fraction in the coma of 3I, we need to choose a model for the non-ice component (amorphous carbon is a typical choice for Solar System comets, whether the dust grains are porous), a physical state for the ice in the coma (physically separate or mixed into dust grains, crystalline vs. amorphous, etc.), and then a scattering mode for each kind of grains (e.g., using \citet{1908AnP...330..377M} theory or \citet{2012tres.book.....H} theory). We tested each of the combinations of models and optical constants used in \citet{2025PSJ.....6..119K}, which were successfully validated on comets in stable activity and undergoing outbursts. Our approach was to model the NIR spectrum on its own, then take those models and see if they could reproduce the spectral behavior seen at visible wavelengths. 

Most models based on the composition and typical grain size distribution of Solar system comets could not reproduce the amount of spectral curvature seen in the near-infrared spectrum, let alone the steeper slopes seen at shorter wavelengths. The only approach which could utilize amorphous carbon and water ice to fit the near-infrared spectrum was to assume a very steep (dominated by small grains) power law distribution of uniformly porous carbon dust grains following the modeling approach of \citet{2018ApJ...862L..16P}. Other styles of models with larger dust grains or intimately mixed dust-and-ice could not reproduce the spectrum at all. The retrieved power-law size slopes $\alpha$ range ($3\sigma$) from 5.5 to 8.8, the lower end of which is on the high end of what is seen in Solar System comets. The very steep power-law  distribution would suggest an unreasonably high abundance of very small coma grains, and may indicate that the assumptions and/or optical constants used in this model, despite the quality of the fit, are only partially right. The three model assumptions most likely to be incomplete are that the comet has a more complicated dust size distribution (e.g., perhaps consists of multiple power laws) than what we've modeled here, the dust grains have porosity or surface texture more complicated than we have modeled here, or the comet has dust with a different composition than most Solar System comets (or at least cannot be modeled using typical techniques). We return to each of these possibilities in the next Section. Regardless, small dust particles of the size discussed here would be highly susceptible to solar radiation pressure, and their presence can therefore be tested through deep imaging observations as the object moves away from opposition. 

We can use this small grain dominated model, plotted in Figure \ref{fig:spectra}, to retrieve a $3\sigma$ upper limit on the areal fraction of pure water ice to be $<7\%$ using the optical constants of carbon from \citet{1991ApJ...377..526R} and crystalline water ice from \citet{2008Icar..197..307M}. This model assumes physically separated dust and ice, so as long as that is true and the dust is approximately the same albedo as Solar System comet dust, this result should be robust. If there is significant ice in 3I's coma, it is not pure small ice grains like that seen in many Solar System comets which is often interpreted \citep{2007Icar..190..284S} as original material that dates from the time of their formation.

\section{Discussion} \label{sec:disc}
\subsection{Ice and Dust}

Although the D-type asteroids and comet nuclei in the Solar system have relatively red and linear visible spectra similar to what we and others see for 3I, their spectra do not display spectral curvature to this extent in the near-infrared. 2I/Borisov showed a similarly-sloped linear visible spectrum \citep{2020MNRAS.495.2053D} but showed a less red but still linear slope in the near-infrared \citep{2020A&A...634L...6Y}. 

We found that the dust distribution required to model the near-infrared spectrum of 3I and not get the visible reflectance completely wrong is anomalously steep and dominated by small grains compared to the comae of Solar System comets. The modeling challenges encountered might indicate something about 3I itself. It may be that 3I's grains are somewhat different in composition than those of typical Solar System comets and thus amorphous carbon is a poor match to them. Later observations at the IRTF reported in \citet{2025arXiv250714916Y} which showed similarly high amounts of spectral curvature were modeled with a combination of dust grains with compositions like that of the meteorite Tagish Lake and pure water ice, which would also be distinct from typical Solar System comet comae. It is also possible that 3I is only weakly active, such that some of our reflectance spectrum is actually reflected light from the nucleus, and thus not modeled appropriately. If 3I really did brighten by half a magnitude or so around discovery \citep{2025arXiv250702757S}, the dust around it at present may not be well described by a single power law.  For 2I/Borisov, its dust grains might have been quite large, unlike many Solar System comets \citep{2021NatAs...5..586Y}, so more complex spectral models will certainly be interesting to apply in the coming months. As 3I/ATLAS is observed at longer wavelengths, larger phase angles, and over a wider time span, these possibilities will become easier to assess.

\subsection{Comparison to 2I/Borisov}
Both of the (unambiguous) Interstellar Comets seem to lack the coma ice seen in many Solar System comets -- particularly at the heliocentric distance at which we observed 3I. Our reflectance spectra showed no obvious signs of absorption features, and our spectral modeling constrained the areal fraction of water ice in the object's coma to be approximately $<7\%$. If the grains in the coma were large and had mixed dust-and-ice compositions, somewhat more ice could be hidden in our data without showing any absorption features. Observations taken approximately ten days after ours were reported in \citet{2025arXiv250714916Y} to show some evidence of water ice, but more work will be needed to disentangle how much is variabilty in the comet on short timescales, how much is due to changes in the comet's activity as it approached the sun, and how much is due to differences in our modeling approaches. Borisov \citep{2020AJ....159...77Y} and ATLAS have both shown cometary activity at distances in which water sublimation is inefficient, suggesting that their activity must be powered by the sublimation of a non-water substance like CO or CO$_2$ (see, e.g., \citealt{2020NatAs...4..867B,2020NatAs...4..861C} for Borisov). This would imply that they haven't been warmed sufficiently to lose all of these more volatile species. These objects could still retain much of the water ice that they formed with -- we just don't see it in their comae. The distant activity and coma composition of comets is a critical way to begin to assess their thermal states and interior structures, and time will tell if the behavior and appearance of Borisov and ATLAS at large heliocentric distances are hints that Interstellar Comets store their ice differently than our own comets do. Until a more complete picture of 3I/ATLAS's volatile inventory begins to form -- as will likely come from observing it when it is at warmer temperatures near perihelion -- it seems challenging to comment on the likelihood of \citet{2025arXiv250705318H}'s suggestion that 3I may be intrinsically water rich based on its dynamics. If the object's water ice were stored in large, mixed grains as opposed to the small pure grains proposed to be nearly-original in Solar System comets (\citealt{2007Icar..190..284S}), its coma could potentially be water rich -- this would just make it distinct from many Solar System comets.

\section{Summary}
We observed the Interstellar Comet 3I/ATLAS with the SpeX, 'Opihi, and MORIS on the NASA Infrared Telescope Facility (IRTF) on 2025 July 3 and 4 to constrain its visible and near-infrared reflectivity, hunt for signatures of water ice in its coma, measure its lightcurve, and compare its properties to other local and interstellar comets. These observations occurred within days of its discovery and thus should be comparable to other observations in its discovery epoch as well as a comparison point for later near-infared studies of the comet.

3I/ATLAS's lightcurve measured with 'Opihi on July 3 was low-amplitude and nearly consistent with flat. While the absence of light curve variability may indicate that 3I's nucleus has a roughly circular shape in projection (and thus a low lightcurve amplitude), it is also possible that 3I is surrounded by enough dust that any nuclear signal would be challenging to identify conclusively. That said, longer rotation periods like those suggested in \citet{2025arXiv250712922D} might be reconcilable with our data. Through `Opihi observations on July 4, we measure an averaged $g'-i'$ color of $0.98\pm0.03$ which is within $1\sigma$ of the red visible reflectivity measured by \citet{2025arXiv250702757S, 2025arXiv250705226O, 2025arXiv250712922D, 2025arXiv250705252B, 2025RNAAS...9..194B}. Given 3I's apparent activity, and non-D-type spectral features seen in the near-infrared, comparing 3I/ATLAS and solid surfaces of similar slopes (including 1I/'Oumuamua) may not be relevant.

Our NIR spectrum taken with SpeX also shows a red and linear reflectance spectrum like the D-type asteroids and comet nuclei, but only below 0.9 $\mu{m}$. At longer wavelengths, the spectrum becomes less red, then neutral (or possibly even slightly blue). No obvious absorption signatures, including those of water ice, are seen. The absence of absorption features is similar to 2I/Borisov \citep{2020A&A...634L...6Y}, but the near-infrared spectra of the two interstellar comets show different NIR slopes and curvatures. The high amount of spectral curvature observed was difficult to reproduce with physically plausible spectral models, assuming it has grain compositions like Solar System comets and a single power law grain size distribution. Our most successful model retrieved an upper limit ($3\sigma$) on the areal fraction of water ice in the coma of 3I/ATLAS of $<7\%$, but combined dust and ice grains as opposed to physically separate populations of dust grains and ice grains may be able to push this number slightly higher. More work will be needed to identify 3I's total volatile inventory to assess \citet{2025arXiv250705318H}'s suggestion that it might be intrinsically water ice rich based on its dynamics, especially in light of \citet{2025arXiv250714916Y}'s later potential detection of icy grains in 3I/ATLAS when it was brighter and somewhat more active. The distant activity of 2I and 3I implies they have some supervolatile ices left to sublimate, and thus that they have not presumably lost all of their less-volatile water ice. However, neither of the Interstellar Comets thus far have shown water ice in their comae despite clear activity beyond where ice sublimates easily, so perhaps these icy visitors store or release their ice(s) somewhat differently than many Solar System comets do.

\begin{acknowledgments}
We thank the staff of the NASA Infrared Telescope Facility for their assistance in the planning, acquisition, and reduction of these observations on such tremendously short notice. The IRTF is operated by the University of Hawaii unde contract to NASA 80HQTR24DA010. We are thankful in particular to the Telescope Operator Greg Engh for significant assistance during the observing run. We wish to acknowledge the significant cultural role that Mauna Kea has always had within the indigenous Hawaiian community. We are very fortunate to have the opportunity to conduct observations from this mountain.

\end{acknowledgments}

\begin{contribution}
All authors contributed to discussions surrounding observational planning, interpretation of results, and to the drafting and editing of this manuscript. T.K. led manuscript writing, modeling and analysis of 3I/ATLAS's spectrum, and preparation of figures. C.C., L.M. and J.E. led observations at the IRTF and reduction of the spectral data with assistance from M.C. and V.R. J.M.B. and B.N.L.S. led analysis of the photometric data, including both colors and lightcurves. M.F. assisted in the interpretation of spectral modeling results.

\end{contribution}

\facilities{IRTF(SpeX,  MORIS, `Opihi)}

\software{\textit{Spextool, \citep{2004PASP..116..362C}}}


\bibliographystyle{aasjournalv7}



\end{document}